\documentclass{aastex}
\usepackage{psfig}

\shorttitle{XMM-Newton RGS Observation of NGC 4636}
\shortauthors{Xu et al.}

\begin{document}

\title{\vspace{-1cm} High Resolution Observations of the Elliptical Galaxy NGC 4636 with the Reflection Grating Spectrometer On-Board XMM-Newton}
\author{
{\sc H.} {\sc Xu}$^{1,2}$, {\sc S. M.} {\sc Kahn}$^{1}$, {\sc J. R.} {\sc
Peterson}$^1$, {\sc E.} {\sc Behar}$^1$, \\
{F. B. S.} {\sc Paerels}$^1$, {\sc R. F.} {\sc Mushotzky}$^{3}$,
{\sc J. G.} {\sc Jernigan}$^{4}$, \\
and {\sc K.} {\sc Makishima}$^{5}$ \\
{\small\it $^{1}$ Columbia Astrophysics Laboratory, 550 W 120th Street, New York,
NY 10027, USA} \\
{\small\it $^{2}$ Department of Applied Physics, Shanghai Jiao Tong University,} \\{\small\it 1954 Huashan Road, Shanghai 200030, PRC} \\
{\small\it $^{3}$ NASA/GSFC, Code 662, Greenbelt, MD 20771, USA}\\
{\small\it $^{4}$ Space Sciences Laboratory, University of California,
Berkeley, CA  94720, USA} \\
{\small\it $^{5}$Department of Physics, University of Tokyo, 7-3-1 Hongo,
Bunkyo-ku, Tokyo 113-0033, Japan}\\}

\begin{abstract}

We present the first high spectral resolution X-ray observation of the giant
elliptical galaxy NGC 4636, obtained with the Reflection Grating Spectrometer
on-board the XMM-Newton Observatory.  The resulting spectrum contains a
wealth of emission lines from various charge states of oxygen, neon,
magnesium, and iron.  Examination of the cross-dispersion profiles of several
of these lines provides clear, unambiguous evidence of resonance scattering
by the highest oscillator strength lines, as well as a weak temperature
gradient in the inner regions of the interstellar medium.  We invoke a
sophisticated new Monte Carlo technique which allows us to properly account
for these effects in performing quantitative fits to the spectrum. Our
spectral fits are not subject to many of the systematics that have plagued
earlier investigations. The derived metal abundances are higher than
have been inferred from prior, lower spectral resolution observations
of this source (Awaki et al. 1994), 
but are still incompatible with conventional chemical
enrichment models of elliptical galaxies. In addition,
our data are incompatible with standard cooling flow models for this system
- our derived upper limit to the mass deposition rate is below the predicted 
value by a factor of 3--5.

\end{abstract}

\keywords{galaxies: individual (NGC 4636)--galaxies: ISM--X-rays: galaxies}

\section{Introduction}

Most elliptical and  other early-type galaxies possess a
hot and diffuse interstellar medium (ISM) that extends
out to several tens of kpc. The ISM temperature ranges from
0.5 keV to 1 keV (e.g., Matsumoto et al. 1997) so that most of its
emission is radiated in the soft X-ray band. Detailed analysis
of the X-ray spectra of these systems, which are dominated by a
wealth of characteristic emission
lines, provides a robust tool to understand the metal enrichment
processes in stars and interstellar gas, and possibly to
constrain the supernova history of the
parent galaxies.

However, in all previous investigations, the spectral resolution
and/or sensitivity of the instruments used were insufficient to cleanly
address these issues. Hundreds of emission lines,
particularly a forest of iron-L lines, were completely blended
in the observed spectra. Therefore, simple model fits to the X-ray
data do not yield unambiguous determinations
of the gas density and temperature, the metal abundances, or
their spatial distributions.

In this paper, we report the first high spectral resolution
X-ray observation of the giant elliptical galaxy NGC 4636,
performed with the Reflection Grating Spectrometer (RGS)
(den Herder et al. 2001) on board the XMM-Newton Observatory.
At $z=0.003129$ (Smith et al. 2000), NGC 4636 is located in
the skirt region of the Virgo cluster. Global fitting of
ASCA SIS spectra
(Matsumoto et al. 1997) yielded a mean temperature for the
interstellar gas in this galaxy of $0.76\pm0.01$ keV and
an average metal abundance of $0.31^{+0.04}_{-0.03}$ solar.
The $0.5-4.5$ keV luminosity was found to be
$5.7\times10^{41}$ erg s$^{-1}$
which makes it one of the brightest elliptical galaxies in
the X-ray band.

In \S2, we describe our observations and basic data
reduction procedures. The detailed analysis of the RGS
spectrum is presented in \S3.  We demonstrate that the observed
spectral and spatial distributions mandate the presence of a weak
temperature gradient in the ISM of NGC 4636, as well as careful
consideration of the importance of resonance line scattering
in interpreting the measured line intensities.  We invoke
a novel modeling and fitting procedure that allows us to constrain
the relevant plasma parameters with these effects included.
Our spectral fits indicate higher metal abundances than
have been inferred from prior, lower spectral resolution observations
of this source, but are still discrepant with expectations from
conventional chemical enrichment models.  In addition, we place
tight upper limits on the contribution
of low temperature gas, in contradiction with the predictions of
standard cooling-flow models for the ISM of this galaxy.  In
\S4, we summarize and discuss the implications of these results.
Throughout the paper, we assume a distance to NGC 4636 of 15.0 Mpc
(Ferrarese et al. 2000).
Unless stated otherwise, the wavelengths of the emission lines
are expressed in a reference frame at rest in the source.
The solar abundance ratios are taken from Anders and Grevesse (1988).

\section{Observations and Data Reduction}

NGC 4636 was observed with XMM-Newton in revolutions 0109
(July 13--14, 2000) and 0197 (January 5, 2001). During
revolution 0109, a historically strong solar flare event
occurred so that the European Photon Imaging Camera (EPIC)
detectors were all switched off, and the RGS spectra were 
heavily contaminated with particle-induced background. In 
this initial report, we therefore concentrate on the RGS 
data obtained during the second observation. The total
exposure is approximately 64 ks. We processed the data with
the Science Analysis System (SAS) for event reconstruction,
CCD energy correction, and angular coordinate mapping.
Approximately $3.5\times10^{4}$ photons are contained in
the spectrum after rejecting bad events.  Given the spatial
extent of the source, the effective spectral resolution
of the RGS is $\Delta\lambda \sim 0.2$ \AA\/
over most of the band.

\section{Spectral Analysis}
\label{sec:specfit}

\subsection{Line Identification}
\label{sec:lines}

The raw RGS spectra, which are uncorrected for the response
of the instrument and the redshift of the source, are plotted
as a function of wavelength in Figure 1. The first- and
second-order spectra obtained by both RGS1 and RGS2 have
been added together in constructing this figure. We have not
removed particle or detector background.
The spectra are dominated by the $2p-3d$ lines of Fe
XVII-XXI ($12-15$ \AA\/), the $2p-3s$ lines of Fe XVII-XVIII
($16-17.1$ \AA\/), and the Ly$\alpha$ lines of O VIII
(19.0 \AA\/) and Ne X (12.1 \AA\/). The blended Helium-like
Mg XI triplet ($\sim9.2$ \AA\/) can also be clearly identified.
In addition, there are weak Ly$\alpha$ lines of N VII
(24.8 \AA\/) and Mg XII (8.4 \AA\/), and the O VII triplet at
$21.6-22.1$ \AA\/. Because the O VIII lines are much stronger
than the O VII lines, and the Fe XVIII-XIX $2p-3d$ lines are
weaker than the Fe XVII $2p-3d$ lines, there must be very little
contribution from gas with $kT<500$ eV or $kT>700$ eV.
The ratios of Mg XII line intensities to those of Mg XI are
consistent with this range, and in fact indicate a temperature of
approximately 600 eV.

Measured incident fluxes of those lines which can
be well isolated are given in Table 1.
These have been obtained by integrating over the cross-dispersion
direction on the RGS detectors, subtracting the relevant background,
and dividing by both the RGS effective area specific to these extractions and
by the effective exposure times.  Of particular interest are the
relative intensities of the brightest Fe XVII lines
at $15.01$ \AA, $15.26$ \AA, $16.78$ \AA, and those
at $17.05$ and $17.10$ \AA\/, which are blended.
Ratios of these lines are listed in Table 2,
along with the analogous quantities deduced from Chandra HETG observations
of Capella (which has a very similar temperature to what we infer
for NGC 4636; Behar et al. 2001), and the predicted line ratios at 
this temperature given by the APEC model for a plasma in collisional 
equilibrium (Smith et al. 2001).  As can be seen, the NGC 4636 Fe XVII 
line ratios are in reasonable agreement with those measured for Capella,
but disagree more significantly
with the APEC predictions.  We find similar effects for Fe XVIII.  In
both cases, the $2p - 3d$ lines are too weak relative to the $2p - 3s$
lines.  The origin of this discrepancy is still unclear,
but, since it is also observed in the Capella spectrum,
it appears to possibly be an artifact of uncertainties in the 
atomic excitation rates for these lines 
(Brown et al. 1998; Laming et al. 2000; but see also Brown et al. 2001), 
and not of the astrophysical conditions in the NGC
4636 ISM. We therefore empirically correct the intraseries Fe L-shell line
ratios in our subsequent modeling, using Capella as the calibrator.
Following Behar et al. (2001), we choose to
normalize the relative line
intensities to the intensity of the 15.26 \AA\/ line for Fe XVII
and its spectroscopic equivalent for Fe XVIII.  This choice gives the
best agreement with the remainder of the L-shell complex for each ion,
particularly the higher series lines.

\subsection{Cross-Dispersion Profiles}
\label{sec:cross}

The RGS is a slitless, nearly stigmatic spectrometer, which means we can place
some constraints on the spatial dependence of the various emission lines,
by examining the profile of the emission line fluxes and their ratios
in the cross-dispersion direction.  We first look at the ratio of the two
$2p - 3s$ lines of Fe XVII at 17.1 \AA\/ to that of the $2p - 3d$ line
of Fe XVII at 15.0 \AA\/, which is plotted in Figure 2a.  As can be
seen, this ratio is significantly peaked toward the center of the gaseous
halo of the galaxy.  Since these lines originate from the same charge
state of a single element,
the observed spatial dependence cannot be due to gradients in either the
temperature or elemental abundances within the ISM.  Instead, it is
most likely produced by resonance line scattering of the 15.0 \AA\/
photons. The oscillator strength for the $2p - 3d$ transition is substantially
higher than that of the $2p - 3s$ lines.  For the inferred densities
and temperatures in the core of the NGC 4636 ISM,
the 15.0 \AA\/ line optical depth is greater than unity,
while the 17.1 \AA\/ blend optical depth is negligible.
The 15.0 \AA\/ photons typically scatter one or more times before
exiting the medium, flattening the intensity profile of that line
in comparison to the others.  Hence, accurate treatment
of resonance line scattering is essential to correctly modeling the
extracted spectrum. 
\footnote{In deriving the line intensities given in Tables 1 and 2,
we used an extraction region of $2'$ in the cross-direction. Since
this is much larger than the angular scale of the observed variations,
resonance scattering does not affect these integrated line intensities.}

In Figure 2b, we plot the ratio of the 17.1 \AA\/ line of Fe XVII to the
16.0 \AA\/ line of Fe XVIII.  Here again, it is centrally
peaked.  However, these are both $2p - 3s$ transitions, and neither has
sufficient optical depth for the profile to be affected by resonance scattering.
In this case, the variation in the ratio must be associated with a
temperature gradient, which makes the Fe XVII and
Fe XVIII ion fractions a function of radius.  The temperature must
drop by $\sim 20-30$ percent from the outer regions of the halo
toward the core.

Finally, in Figure 2c, we plot the ratio of the Fe XVII 17.1 \AA\/ blend
to the Ly$\alpha$ line of O VIII at 19.0 \AA\/.  These lines have similar,
but not exactly identical emissivity dependences on temperature.  As can
be seen, there is only a mild spatial dependence, of the same order
as expected for the inferred temperature gradient.  The implication is that
there is not a severe relative abundance gradient in the halo.  We also
find no strong evidence of abundance gradients by looking at the profiles 
of line ratios involving the other elements.

\subsection{Quantitative Spectral Fitting}
\label{sec:fitting}

As we have shown, proper modeling of the gas distribution in the ISM
of NGC 4636 requires careful treatment of both temperature gradients
and resonance line transfer.  In addition, since this is
an extended source, the line spread function of the RGS instrument
is dependent on the spatial distribution of the emission.  All of these
factors require that the spatial and spectral distributions be
modeled jointly.  Standard spectral fitting procedures cannot be used
for this purpose.

We have developed a new approach to spectral fitting of extended
sources which relies on Monte Carlo methods (Peterson, Jernigan, \&
Kahn 2001).  Assuming given 3-dimensional
spatial distributions for the relevant plasma
parameters, Monte Carlo photons are generated within the medium at positions
and wavelengths weighted by the emissivity model.  These photons
are then scattered
in both frequency and space based on line scattering probabilities,
evaluated assuming velocity widths appropriate to maxwellian doppler
broadening at the local gas temperature.  At each scatter, the photon is
given a new trajectory and frequency
until it escapes from the model medium.  After
projection onto the sky, the photons are propagated through an
instrument Monte Carlo simulator to predict the eventual detected position
in both the dispersion and cross-dispersion coordinates, and the CCD
charge content of the event.  A separate Monte Carlo algorithm,
calibrated on deep observations of the Lockman Hole, is used to generate
an appropriate sample of background events.  The final simulated data
set is subjected to the identical set of extractions as are applied to the
measured data.  The observed and simulated data are compared by means
of a $\chi^{2}$ statistic in each of the various dimensions.
This process is repeated by iterating on the parameters characterizing
the spatial and spectral distributions until an acceptable fit is obtained.
In order to maintain high accuracy in the simulation, the number of simulated
photons is chosen to be ten times higher than that of the observed data
set.

We assume a $\beta$ profile (Cavaliere \& Fusco-Femiano 1976) to 
approximate the spatial distribution of the electron density within 
the central 2':
\begin{equation}
\label{eq:betagas}
n(R)=n_{0}
        \left[
        1+(R/R_{c})^{2}
        \right]^{-3\beta/2}.
\end{equation}
The core radius, $R_{c}$, and the $\beta$ parameter were determined
from fits derived from an available Chandra
ACIS observation (Mushotzky et al. 2001), and were found to be 9" and
0.45 respectively.  The central density, $n_{0}$, is left as a
free parameter.  To accommodate the observed temperature gradient,
we assume a temperature distribution of the form:
\begin{equation}
\begin{array}{lll}
T(R) & = & \left\{ \begin{array}{ll}
         T_{\rm min}+(T_{\rm max}-T_{\rm min})(\frac{R}{R_{c}})^{\alpha} &  
         \mbox {if $R\leq R_{c}$} \\
         T_{\rm max} & 
         \mbox {if $R>R_{c}$}            
\end{array}
\right.
\end{array}
\end{equation}
with $\alpha$, $T_{\rm min}$, and $T_{\rm max}$ left as free parameters.

To model the gas emissivity, we use the APEC model, with the Fe L-shell
$2p - 3s$ to $2p - 3d$ ratios adjusted to match the Capella values, as
discussed in \S3.1.  The abundances of N, O, Ne, Mg, and Fe (which account
for all of the observed features in the spectrum) are left as free
parameters.  The best fit values of all of these parameters are listed in
Table 3.  Here, the value of $\chi^{2}$ quoted is for the comparison
to the spectrum displayed in Figure 1, where the model has been overlayed.
As can be seen, the fit is quite good.  There are only small residuals
in the vicinities of the brightest lines, possibly indicating that
the assumed $\beta$-profile may still be too much of an over-simplication.
The predicted cross-dispersion
profiles for this same model are also plotted in the various panels of
Figure 2.  In general, there is excellent agreement with the data.  We have
not allowed for any abundance gradients in the model, and none appear
to be required.  The best-fit temperature distribution increases gradually
from $kT = 0.55$ keV at $R = 0$ to $kT = 0.70$ keV at $R = 42"$.
With this temperature
profile, the observed temperature-sensitive $I$(17.1\AA\/)/$I$(16.0\AA\/)
ratio is fitted very well (Figure 2b).

Note that the $I$(17.1\AA\/)/$I$(15.0\AA\/) ratio is also well-described
by the model (Figure 2a).  This validates our conjecture that the observed
variation in this ratio is associated with resonance scattering of the
15.0 \AA\/ photons.  We emphasize that
the gas density was not independently
adjusted to fit this variation, i.e. the same density profile necessary
to describe the observed intensity distribution 
(at the same assumed distance)
also quantitatively accounts
for the observed resonance scattering effects. In addition, our derived
value for the central gas density is in good agreement with that found by
Loewenstein et al. (2001) for the bremsstrahlung continuum in the Chandra
ACIS spectrum.

In these spectral fits, no extra absorption beyond that associated
with the Galactic column,
$N_{\rm H}=1.87\times 10^{20}$ cm$^{-2}$ (Murphy et al. 1996), 
is required. With our best-fit parameters, we find that the total
X-ray luminosity within 1' is $1.87\pm0.19\times10^{41}$ erg s$^{-1}$.
This agrees well with the ROSAT (Matsushita 2001) and the
recent Chandra (Loewenstein et al. 2001) measured values.

We have not attempted to develop and fit
a full spatial-spectral cooling flow model for NGC 4636.
However, important constraints on cooling flow models can be derived from
the observed luminosities in the Fe XVII and O VII emission lines.  The
multiphase, isobaric cooling flow model of Johnstone et al. (1992) predicts
a constant luminosity per temperature interval that is proportional to
the mass deposition rate of cooling gas.  This radiation is in addition
to that emitted by the ambient gas, which has not yet cooled.  Individual
spectral line fluxes can thus be used to derive upper limits to the mass
deposition rate, applicable within particular temperature ranges of
sensitivity.

Assuming the $\beta$-profile described above, our observed
Fe XVII line intensities imply an upper limit to the mass deposition rate
of 0.21 $M_{\odot}$ yr$^{-1}$.  Assuming that iron L-shell emission
dominates the cooling of the gas, which is true in this temperature range,
this derived limit on the mass deposition rate is relatively insensitive
to the iron abundance.  That is not true for oxygen.  However, we can
fix the O/Fe abundance ratio using the relative intensity of the
O VIII Ly$\alpha$ line.  With that constraint, the measured
O VII line intensities near 22 \AA\/ imply an upper limit on the cooling
flow mass deposition rate of 0.30 $M_{\odot}$ yr$^{-1}$.  Note that these
limits are a factor 3--5 below the inferred theoretical 
mass deposition rates for the halo of this galaxy derived from imaging 
observations (e.g., Bertin \& Toniazzo 1995). They are also
significantly below the value, $0.43\pm0.06$ $M_{\odot}$ yr$^{-1}$,
inferred from O VI observations in the UV band by Bregman et al. (2001).
Similar discrepancies between mass deposition rates inferred from imaging 
data and those allowed by the RGS spectra have been found for massive cooling
flow in clusters of galaxies (Peterson et al. 2001; Kaastra et al. 2001;
Tamura et al. 2001) and had even been suggested by the original ASCA
data (cf. Makishima et al. 2001).

\section{Discussion}

Giant elliptical galaxies are the most massive and the oldest stellar
systems in the universe.  Analyses of their stellar spectra, and their
evolution with time has indicated that most of the stars formed at
very high redshift.  There is little evidence of any significant
star formation within the last 5 Gyrs.  The hot X-ray emitting ISM in
these systems is the repository of the total mass loss from stellar
winds, planetary nebulae, and supernovae.  As such, its metal abundances
are very sensitive to both the abundances in the stars themselves, and
to the Type Ia supernova rate.

Previous analyses of X-ray emission from ellipticals (cf., Arimoto
et al. 1997) presented some fundamental challenges to our 
understanding of the formation and evolution of these systems.
Given that the stars are believed to be supermetal rich, and that
supernovae only add metals to the interstellar gas, it was expected 
that the X-ray spectra would yield metal abundances well in excess 
of solar. The general picture was that the low-Z elements should be 
strongly skewed to those of massive supernovae (Type IIs), while Fe 
should show a particular enhancement, as high as 2 - 5 times solar, 
due to the contributions of Type Ia supernovae.

The first quantitative test of these predictions came with ASCA, which
provided moderate resolution X-ray spectra, suitable for identifying
broad emission ``humps" associated with each of the major elemental
constituents.  Contrary to expectations, fits to ASCA spectra invariably
yielded significantly subsolar abundances (Matsumoto et al. 1997).  In the 
intervening years, a variety of suggestions have been offered to explain 
this discrepancy: Problems with the iron L-shell atomic physics in the 
available spectral codes (Arimoto et al. 1997; Matsushita, Ohashi \&\/ Makishima 2000); 
multi-temperature distributions in the emitting gas (Buote 1999); abundance 
gradients and metallicity dependent supernova rates (Finoguenov \& Jones 2000); 
excess helium abundance (Drake 1998).  The multitude of possibilities arises 
from the fact that the available ASCA data did not have sufficient spatial 
or spectral resolution to constrain the situation in any real detail.  
Abundance information could only be derived by global fits to the entire 
spectrum, which are known to produce misleading results if the assumptions 
invoked are invalid.  In fact, as shown in Matsushita et al. (1997), the iron 
abundance can be increased to $\sim 1$ solar if systematic errors are allowed 
at 20\%\/ level in the spectral fittings to ASCA data.

It is worth noting that the abundance patterns of the stars in
giant ellipticals are also subject to uncertainties.  In particular,
Trager et al. (2000) and Kobayashi and Arimoto (1999) have shown that
one can really only reliably determine the relative abundances of the
alpha elements, and the Fe/alpha ratio.  It appears that there is often
a steep alpha element abundance gradient in the inner regions of the 
galaxy, without an accompanying gradient in the iron abundance. Such 
effects call into question the simple chemical enrichment models that 
have previously been developed for these systems.

The XMM-Newton RGS observations of NGC 4636 presented here provide the most
accurate and unambiguous abundance determinations to date for the ISM
of a giant elliptical galaxy.  We have shown that there is a weak, but
measurable, temperature gradient in the inner regions of the ISM.  In
addition, we have shown that resonance scattering of the high oscillator
strength iron lines is non-negligible, consistent with expectations given the
measured central electron density in the gas.  Both of these effects are
very important, and must be properly taken into account in the derivation
of abundance constraints from the X-ray spectrum.  The high statistical
quality of our data, coupled with the high spectral and spatial resolution
that the RGS provides, makes our abundance and temperature constraints
far more robust than any previous determinations.

Nevertheless, our derived values:  O/Ne $\sim0.7$ solar, O/Mg $\sim0.8$ 
solar, O/Fe $\sim$ 0.6 solar, and Fe $\sim$ 0.87 solar, are still difficult 
to reconcile with the chemical enrichment models.  In particular, these 
values do not match the Type Ia and Type II models in Gibson, Lowenstein 
and Mushotzky (1997), nor any linear superposition of these models.  The 
iron abundance has been increased significantly compared with those previous
measurements, but it
is still lower than the theoretical predictions.  One possible explanation
is that iron has been escaping from the ISM, as inferred by the observed large 
scale metal abundance gradient in NGC 4636 (Mushotzky et al. 1994; Matsushita
et al. 1997).  We believe that our results and future investigations 
set a standard to which new models of the evolution of giant ellipticals 
should be compared.

Equally puzzling is the clear lack of a measurable cooling flow in the
ISM of NGC 4636.  It has been well accepted for over fifteen
years (e.g., Nulsen et al. 1984) that the central regions of giant elliptical
galaxies should exhibit strong cooling flows.  For NGC 4636, our inferred
central density indicates a cooling time of less than $10^8$ years.  This is
compatible with the sound crossing time evaluated at the local sound speed,
which means that the gas should be cooling as fast as it possibly can.
There are no obvious sources of heating in this gas - Type Ia supernovae
and stellar winds provide too little energy.
In addition, recent Chandra observations (Lowenstein et al. 2001)
have shown that there is no active nucleus present in this galaxy.  A
previous outburst from a now dormant active nucleus could provide the
required heat (Jones et al. 2001), but it is unclear yet how common
such outbursts might be.

However, as we have shown the weakness of the Fe XVII and O VII lines,
and the absence of gas at temperatures less than 500 eV are in strong 
disagreement to the cooling flow models. There is only a weak temperature 
gradient and no other evidence of multiphase gas.  In particular, the 
higher resolution RGS data do not support the conclusions of Buote \&\/ 
Fabian (1998), which were derived from ASCA observations of this source.

As noted in \S3.3, problems with cooling flow models have also emerged
from XMM-Newton RGS observations of clusters of galaxies.  The issue is
actually more severe for NGC 4636 given the wealth of discrete spectral
lines available in the lower temperature range appropriate to elliptical
halos.  It is conceivable that our data could be compatible with a model
in which some unknown form of rapid cooling occurs, radiating most of
the energy in the ultraviolet.  The fact that the O VI measurements
from FUSE (Bregman et al. 2001) suggest a higher mass deposition rate than
allowed by our O VII upper limits might argue for such a picture.  Turbulent
mixing layers (Begelman \& Fabian 1990) could possibly accomplish this
by entraining the X-ray gas and rapidly mixing it down to temperatures
of order $10^5$ K.  Additionally, the distortion of the emission measure
distribution by adiabatic compression (Nulsen 1998) might conceivably
explain our results given the relatively narrow temperature range we are
sampling.  Regardless of the eventual explanation, this observation does
demonstrate that the discrepancies with cooling flow predictions occur
even on small scales, not only for the massive flows in galaxy clusters.

This work was supported by a grant from NASA to Columbia University for
XMM-Newton mission support and data analysis. We wish to thank other
key members of the RGS team for their advice on instrument calibration
and data analysis issues. H. Xu is also partially supported by the National
Science Foundation of China (Grant No. 19803003) and the Ministry of Science
and Technology of China (Project 973).

\newpage

{\small
\begin{table}[h!]
\caption{Measured line fluxes from the central 2' of NGC 4636}
\label{tbl:flux}
\begin{center}
\begin{small}
\begin{tabular}{llllclcc}
\hline \hline
\multicolumn{1}{c}{$\lambda_{\rm observed}$}&
\multicolumn{1}{c}{$\lambda_{0}^{*}$}&
\multicolumn{1}{c}{Ion}&
\multicolumn{2}{c}{Upper Level}&
\multicolumn{2}{c}{Lower Level}&
\multicolumn{1}{c}{Flux}\\
\multicolumn{1}{c}{(\AA\/)}&
\multicolumn{1}{c}{(\AA\/)}&
&
\multicolumn{1}{c}{Configuration}&
\multicolumn{1}{c}{$J_{U}$}&
\multicolumn{1}{c}{Configuration}&
\multicolumn{1}{c}{$J_{L}$}&
\multicolumn{1}{c}{($10^{-4}$ ph s$^{-1}$ cm$^{-2}$)}\\
\hline
12.1&12.132&Ne$^{9+}$&
 2p$_{3/2}$ &3/2&
 1s         &1/2&
 2.24\\
 &12.137&Ne$^{9+}$&
 2p$_{1/2}$ &1/2&
 1s         &1/2&
 \\
\hline
14.2&14.208    &Fe$^{17+}$&
 2p$_{1/2}$2p$^3_{3/2}$3d$_{3/2}$ &5/2&
 2p$^2_{1/2}$2p$^3_{3/2}$         &3/2&
 1.90\\
 &14.208&Fe$^{17+}$&
 2p$_{1/2}$2p$^3_{3/2}$3d$_{5/2}$ &3/2&
 2p$^2_{1/2}$2p$^3_{3/2}$         &3/2&
 \\
\hline
14.3&14.256&Fe$^{17+}$&
 2p$_{1/2}$2p$^3_{3/2}$3d$_{3/2}$ &1/2&
 2p$^2_{1/2}$2p$^3_{3/2}$         &3/2&
 1.45\\
 &14.267&Fe$^{19+}$&
 2s2p$^{2}_{1/2}$2p$_{3/2}$3s &3/2&
 2s2p$^2_{1/2}$2p$^2_{3/2}$             &5/2&
 \\
\hline
14.4&14.373&Fe$^{17+}$&
 2p$_{1/2}$2p$^3_{3/2}$3d$_{3/2}$ &5/2&
 2p$^2_{1/2}$2p$^3_{3/2}$         &3/2&
 1.85\\
\hline
14.5&14.534&Fe$^{17+}$&
 2p$^2_{1/2}$2p$^2_{3/2}$3d$_{5/2}$ &5/2&
 2p$^2_{1/2}$2p$^3_{3/2}$           &3/2&
 2.25\\
 &14.571&Fe$^{17+}$&
 2p$^2_{1/2}$2p$^2_{3/2}$3d$_{5/2}$ &3/2&
 2p$^2_{1/2}$2p$^3_{3/2}$           &3/2&
 \\
\hline
15.0&15.014&Fe$^{16+}$&
 2p$_{1/2}$2p$^4_{3/2}$3d$_{3/2}$ &1&
 2p$^2_{1/2}$2p$^4_{3/2}$       &0&
 3.56\\
\hline
15.3&15.261&Fe$^{16+}$&
 2p$^2_{1/2}$2p$^3_{3/2}$3d$_{5/2}$ &1&
 2p$^2_{1/2}$2p$^4_{3/2}$         &0&
 1.66\\
\hline
16.1&16.004&Fe$^{17+}$&
 2p$^{2}_{1/2}$2p$^{2}_{3/2}$3s &3/2&
 2p$^{2}_{1/2}$2p$^{3}_{3/2}$   &3/2&
 1.99\\
 &16.071&Fe$^{17+}$&
 2p$^{2}_{1/2}$2p$^{2}_{3/2}$3s &5/2&
 2p$^{2}_{1/2}$2p$^{3}_{3/2}$   &3/2&
 \\
 &16.110&Fe$^{18+}$&
 2s$^{2}$2p$_{1/2}$2p$^{2}_{3/2}$3p$_{1/2}$ &2&
 2s2p$^{2}_{1/2}$2p$^{3}_{3/2}$             &2&
 \\
\hline
16.8&16.780&Fe$^{16+}$&
 2p$_{1/2}$2p$^4_{3/2}$3s &1&
 2p$^2_{1/2}$2p$^4_{3/2}$ &0&
 2.33\\
\hline
17.0&17.051&Fe$^{16+}$&
 2p$^2_{1/2}$2p$^3_{3/2}$3s &1&
 2p$^2_{1/2}$2p$^4_{3/2}$   &0&
 2.77\\
\hline
17.1&17.096&Fe$^{16+}$&
 2p$^2_{1/2}$2p$^3_{3/2}$3s &2&
 2p$^2_{1/2}$2p$^4_{3/2}$   &0&
 2.89\\
\hline
19.0&18.967&O$^{7+}$&
 2p$_{3/2}$ &3/2&
 1s         &1/2&
 1.96\\
 &18.973&O$^{7+}$&
 2p$_{1/2}$ &1/2&
 1s         &1/2&
 \\
\hline
\end{tabular}
\end{small}
\end{center}
\begin{description}
  \begin{footnotesize}
  \setlength{\itemsep}{-1mm}
     \item[$*$] $\lambda_{0}$ is the wavelength measured with EBIT for iron (Brown et al. 1998, 2000), or calculated in HULLAC for other elements (cf. Behar et al. 2001).
  \end{footnotesize}
  \end{description}
\end{table}
}

{\small
\begin{table}[h!]
\caption{Fe XVII line ratios measured in NGC 4636 and Capella}
\label{tbl:ratio}
\begin{center}
\begin{scriptsize}
\begin{tabular}{cccc}
\hline \hline
     &$2p_{3/2}-3s/2p_{1/2}-3s$&$2p_{1/2}-3d_{3/2}/2p_{3/2}-3d_{5/2}$&$2p_{3/2}-3s/2p_{1/2}-3d_{3/2}$\\
&I(17.1)/I(16.8)&I(15.0)/I(15.3)&I(17.1)/I(15.0) \\
\hline
NGC 4636             &$2.30\pm0.18$&$2.31\pm0.18$&$1.40\pm0.13$ \\
Capella$^{*}$        &$2.37\pm0.19$&$2.42\pm0.22$&$1.65\pm0.11$ \\
APEC (600 eV)$^{**}$ &$2.28$&3.49&$1.01$\\
\hline
\end{tabular}
\end{scriptsize}
\end{center}
\begin{description}
  \begin{footnotesize}
  \setlength{\itemsep}{-1mm}
     \item[$*$] From Behar et al. 2001.
     \item[$**$] Predictions of APEC model (Smith et al. 2001).
  \end{footnotesize}
  \end{description}
\end{table}
}

{\small
\begin{table}[h!]
\caption{Best-fit spectral model (errors are 90\%\/ confidence)}
\label{tbl:specfit2}
\begin{center}
\begin{small}
\begin{tabular}{cccccc}
\hline \hline
$kT_{\rm min}$&$kT_{\rm max}$&$\alpha$&$R_{c}$&$n_{0}$&$\chi^{2}_{r}$\\
(keV)&(keV)&&(arcmin)&(cm$^{-3}$)&\\
\hline
$0.53\pm0.03$&$0.71\pm0.02$&$5.8\pm0.3$&$42\pm4$&$0.125\pm0.005$&1.18\\
\hline \hline
\multicolumn{6}{c}{Metal abundances in solar units}\\
Fe&O&Mg&Ne&N&\\
\hline
$0.87\pm0.05$&$0.50\pm0.05$&$0.65\pm0.07$&$0.70\pm0.07$&$1.00\pm0.10$&\\
\hline
\end{tabular}
\end{small}
\end{center}
\end{table}
}

\newpage

{\bf Figure Captions}\\

\noindent
Fig.1--The raw, extracted RGS spectrum of NGC 4636 (black), plotted as
a function of wavelength.  The data from both first and second spectral
orders from RGS1 and RGS2 have been added together in constructing
this histogram.  The best fit model, allowing for a temperature
gradient and accounting for resonance line scattering effects, is
plotted in red.  The residuals to the model fit are plotted below. \\

\noindent
Fig.2--The ratios of various emission lines (as indicated in the panels),
plotted as a function of position in the cross-dispersion direction.  The
data (with appropriate error bars) are plotted as crosses.  The predictions
of the best-fit spatial/spectral model are plotted as diamonds. \\

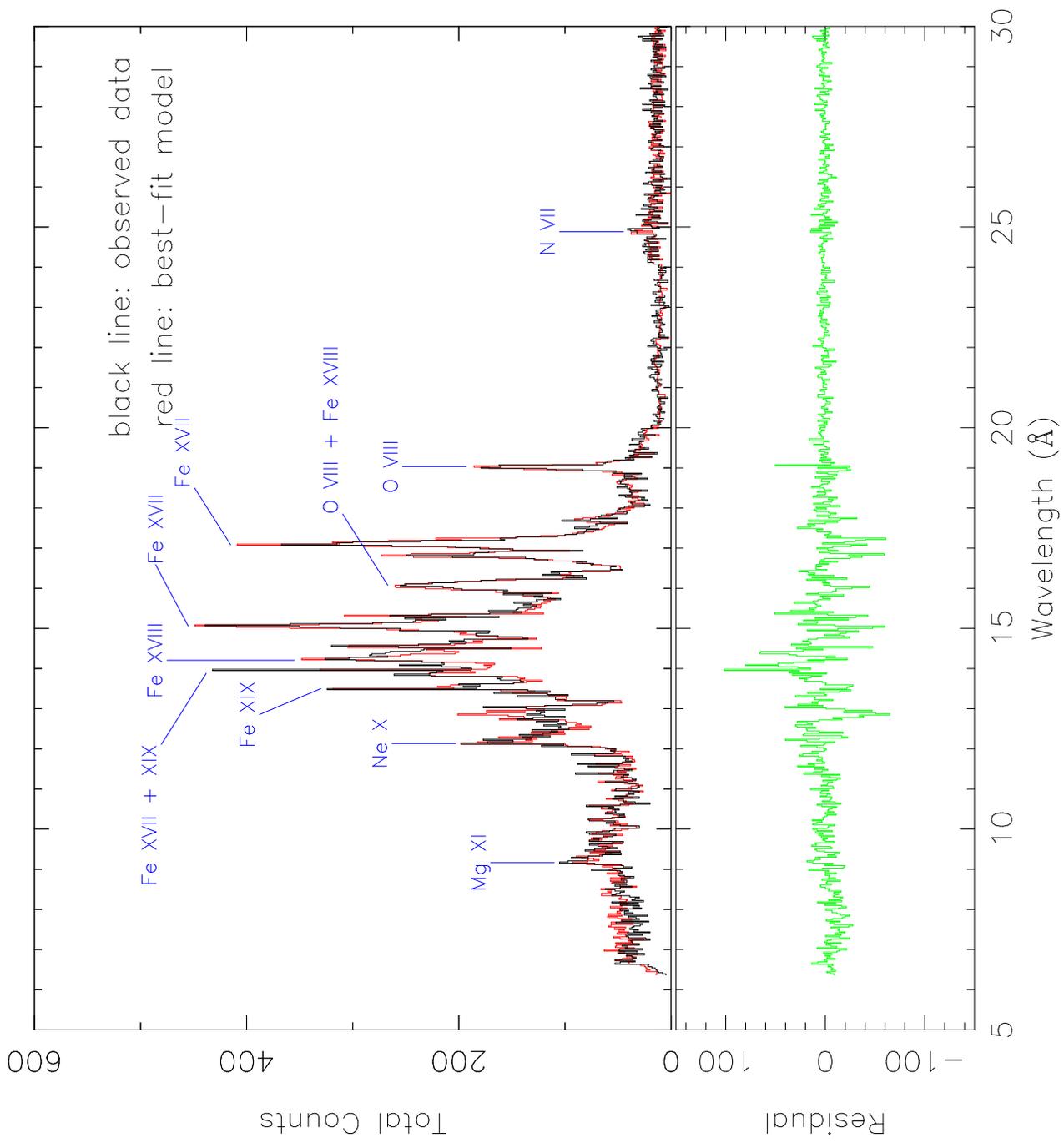
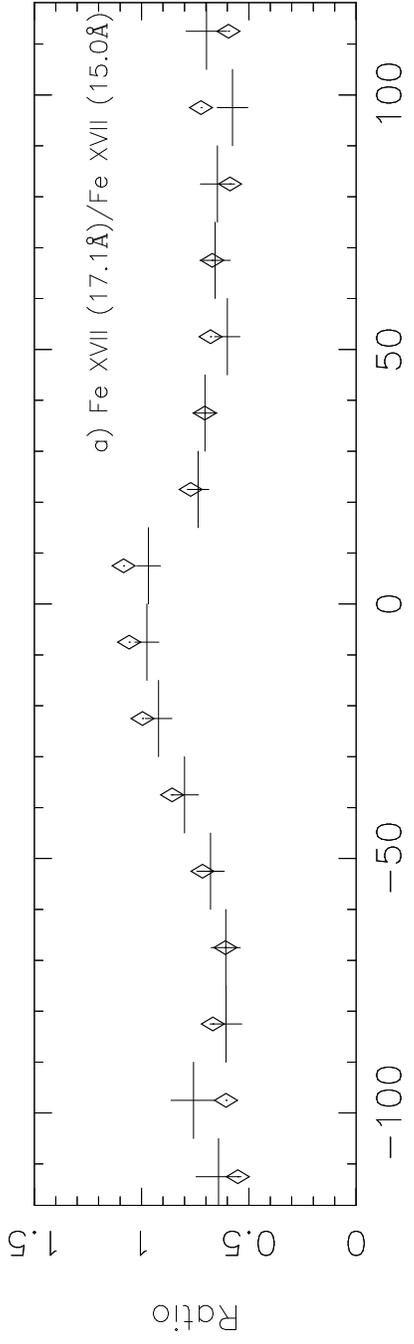
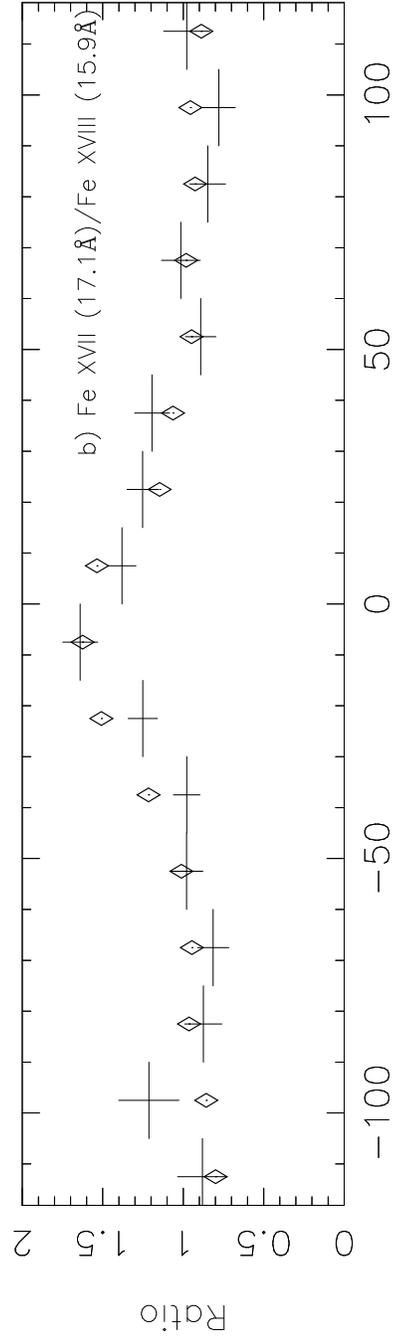
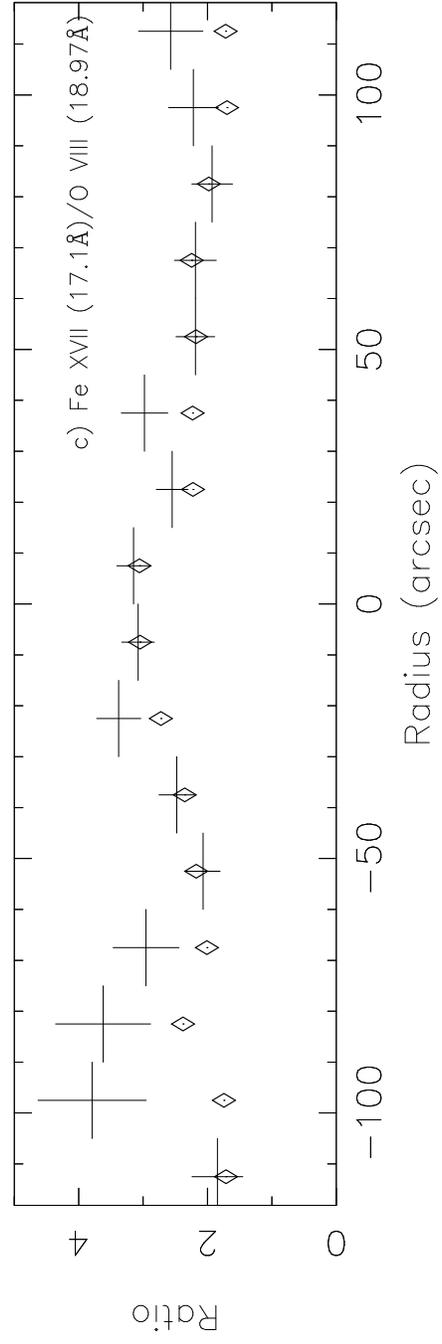
\end{document}